\documentclass[conference]{IEEEtran}
\usepackage{cite}
\usepackage{amsmath,amssymb,amsfonts}
\usepackage{graphicx}
\usepackage{textcomp}
\def\BibTeX{{\rm B\kern-.05em{\sc i\kern-.025em b}\kern-.08em
    T\kern-.1667em\lower.7ex\hbox{E}\kern-.125emX}}

\usepackage{multirow}%
\usepackage{amsthm}%
\usepackage{mathrsfs}%
\usepackage[dvipsnames]{xcolor}
\usepackage{textcomp}%
\usepackage{manyfoot}%
\usepackage{booktabs}%
\usepackage{algorithm}%
\usepackage{algorithmicx}%
\usepackage{algpseudocode}%
\usepackage{listings}%

\usepackage{subcaption}
\usepackage{array, tabularx, bm}
\usepackage{adjustbox}
\usepackage{arydshln}
\usepackage{pifont}
\usepackage{capt-of}
\usepackage{float,wrapfig}
\usepackage{comment}
\usepackage{url}
\usepackage{mathtools, empheq}
\usepackage{enumitem}
\usepackage{colortbl}

\newcommand{\cmark}{\ding{51}} 
\newcommand{\xmark}{\ding{55}} 

\newcolumntype{P}[1]{>{\centering\arraybackslash}p{#1}}

\begin{document}

\title{MoWE-Audio: Multitask AudioLLMs with \\ Mixture of Weak Encoders
}

\author{\IEEEauthorblockN{Wenyu Zhang$^{\dagger}$, Shuo Sun$^{\dagger}$, Bin Wang$^{\dagger}$, Xunlong Zou$^{\dagger}$, Zhuohan Liu$^{\dagger}$, \\ Yingxu He$^{\dagger}$, Geyu Lin$^{\dagger}$, Nancy F. Chen$^{\dagger\star}$, Ai Ti Aw$^{\dagger}$}
\IEEEauthorblockA{\textit{$^{\dagger}$ Institute for Infocomm Research (I$^\text{2}$R), Agency for Science, Technology and Research (A*STAR)} \\
\textit{$^{\star}$ Centre for Frontier AI Research (CFAR), Agency for Science, Technology and Research (A*STAR)}\\
Singapore \\
\{zhang\_wenyu, sun\_shuo, wang\_bin, zou\_xunlong, liu\_zhuohan, \\ he\_yingxu, lin\_geyu, nfychen, aaiti\}@i2r.a-star.edu.sg}
}


\maketitle

\begin{abstract}
The rapid advancements in large language models (LLMs) have significantly enhanced natural language processing capabilities, facilitating the development of AudioLLMs that process and understand speech and audio inputs alongside text. Existing AudioLLMs typically combine a pre-trained audio encoder with a pre-trained LLM, which are subsequently finetuned on specific audio tasks. However, the pre-trained audio encoder has constrained capacity to capture features for new tasks and datasets. To address this, we propose to incorporate mixtures of `weak' encoders (MoWE) into the AudioLLM framework. MoWE supplements a base encoder with a pool of relatively light-weight encoders, selectively activated based on the audio input to enhance feature extraction without significantly increasing model size. Our empirical results demonstrate that MoWE effectively improves multi-task performance, broadening the applicability of AudioLLMs to more diverse audio tasks.
\end{abstract}


\section{Introduction}
\label{sec: introduction}

The transformative advancements in large language models (LLMs) \cite{Achiam2023GPT4TR, Touvron2023Llama2, vicuna2023} have revolutionized the field of natural language processing and artificial intelligence, enabling systems to perform tasks that require a deep understanding of human language. Building upon the foundation of LLMs, there is increasing interest to develop multimodal models to integrate multiple data modalities, such as text, images and audio, into a single end-to-end model \cite{tang2024salmonn, chu2023qwenaudio, deshmukh2023pengi, hu2024wavllm, Das2024SpeechVerse, liu2023llava, dai2024instructblip}. The integration of multiple data modalities allows the model to have more comprehensive understanding of context and allows the users to have more diverse ways to interact with the model.

In this paper, we focus on Audio Large Language Models (AudioLLMs), the class of multimodal LLMs that process and understand speech and audio inputs in conjunction with text. Existing AudioLLMs \cite{he2024meralionaudiollmtechnicalreport, singaudiollm, tang2024salmonn, chu2023qwenaudio,deshmukh2023pengi, Das2024SpeechVerse, wang2023viola, shu2023llasm, rubenstein2023audiopalm, Kong2024AudioFlamingo, ghosh2024GAMA} are capable of executing tasks for both speech and non-speech audio inputs. 
Speech tasks include automatic speech recognition, speech-to-text translation and speech question answering, and non-speech audio tasks include audio captioning, sound event classification and audio question anwering. 
Typically, a pre-trained audio encoder is connected to a pre-trained LLM, and the model is instruction finetuned with curated audio task datasets \cite{he2024meralionaudiollmtechnicalreport, tang2024salmonn, chu2023qwenaudio, deshmukh2023pengi, hu2024wavllm, Das2024SpeechVerse}. The audio encoder is pre-trained with supervised learning \cite{radford2023whisper} or self-supervised learning algorithms \cite{chen2022wavlm, hsu2021hubert, chen2023beats}. Recent AudioLLMs \cite{he2024meralionaudiollmtechnicalreport, tang2024salmonn, chu2023qwenaudio, hu2024wavllm} have utilized the Whisper-large encoder \cite{radford2023whisper}, an advanced model that specializes in speech recognition and speech translation.

Although state-of-the-art pre-trained audio encoders can be utilized, they are typically pre-trained on specific tasks and datasets, and may not have sufficient knowledge and capacity for new tasks and datasets required for general-purpose AudioLLMs. To increase encoder capacity for feature extraction, we propose to incorporate mixtures of `weak' encoders (MoWE) to the AudioLLM framework to supplement the `strong' base encoder. We use the terms `weak' and `strong' encoders to loosely correspond to the concepts of 'weak' and 'strong' learners in machine learning. We define `weak' encoders as those with at least an order of magnitude fewer parameters than the base encoder, and may have lower embedding quality measured by downstream task performance as in Table~\ref{tab: encoder_defn}. This design choice is made so as to not excessively increase model size. 
To constrain the number of active parameters size, we propose routing strategies to selectively activate a subset of weak encoders from an encoder pool for each data sample. Embeddings from the activated encoders are concatenated to be further processed in the AudioLLM pipeline. We demonstrate empirically that MoWE effectively improves multi-task performance.

\begin{table}[b]
\centering

\caption{Comparison between `strong' encoder Whisper-large and relatively light-weight `weak' encoders. Training and validation loss is computed based on the encoder + Llama-3-8B-Instruct experimental setup detailed in Section~\ref{sec: experimental setup}.
\label{tab: encoder_defn}}

\begin{adjustbox}{max width=\columnwidth}
\begin{tabular}{P{2.7cm}P{1.5cm}P{1.7cm}P{1.7cm}}
\toprule[1pt]\midrule[0.3pt]
\textbf{Encoder} & \textbf{\# Param} & \textbf{Train loss ($\downarrow$)} & \textbf{Val. loss ($\downarrow$)}  \\ \midrule
Whisper-large \cite{radford2023whisper}   & 637M  & 0.162 & 0.672\\ \hdashline
Whisper-tiny \cite{radford2023whisper}   & 9M    & 0.401 & 0.749\\
HuBERT-base \cite{hsu2021hubert}    & 95M   & 0.422 & 0.741\\
HuBERT-base-ER \cite{hubert-base-superb-er} & 95M   & 0.407 & 0.733\\
\midrule[0.3pt]\bottomrule[1pt]
\end{tabular}
\end{adjustbox}

\vspace{-4mm}

\end{table}

\section{Related Works}
\label{sec: related works}

\subsection{AudioLLMs}

We focus on the line of work that address audio tasks through an end-to-end pipieline that connects audio encoders to an LLM to generate free-form text responses based on audio inputs. Some models \cite{wang2023viola, shu2023llasm, rubenstein2023audiopalm, Das2024SpeechVerse} target only speech tasks, some \cite{Kong2024AudioFlamingo, ghosh2024GAMA} target only non-speech audio tasks, and others \cite{tang2024salmonn, chu2023qwenaudio, deshmukh2023pengi, gong2023ltuas} address a mix of speech and non-speech audio tasks. The tangential line of work where an LLM acts as an orchestrator to direct data samples to task-appropriate audio foundation models is not considered in this work.

The mainstream architecture employed in multi-modal LLMs, including AudioLLMs \cite{he2024meralionaudiollmtechnicalreport, tang2024salmonn, chu2023qwenaudio, deshmukh2023pengi, hu2024wavllm, Das2024SpeechVerse} and vision-language \cite{liu2023llava, dai2024instructblip} models, consists (1) an encoder to embed the non-language-modality input, (2) a modality adapter and projection to process the resulting embeddings to produce tokens that align with the LLM input space, and (3) an LLM that generates free-form text based on the resulting tokens and textual instruction prompts. For instance, Qwen-Audio \cite{chu2023qwenaudio} connects a Whisper-large-v2 \cite{radford2023whisper} encoder to a Qwen-7B \cite{bai2023qwen} LLM. Some works use two audio encoders to capture the semantic and acoustic information in the audio inputs. SALMONN \cite{tang2024salmonn} connects a Whisper-large-v2 and a Beats \cite{chen2023beats} encoder to Vicuna-13B \cite{vicuna2023}, and WavLLM \cite{hu2024wavllm} connects a Whisper-large-v2 and a WavLM-base \cite{chen2022wavlm} encoder to Llama-2-chat-7B \cite{Touvron2023Llama2}. We incorporate more encoders through mixture of experts routing. Furthermore, we show that even incorporating the same type of encoders can improve performance.

\subsection{Mixture of Experts}

Mixture of experts (MoE) has been extensively explored in LLMs to increase model capacity. In each MoE layer, multiple expert networks are trained, but only a small subset of experts are selectively activated to process each token. MoE is typically applied to feed-forward layers within transformer blocks \cite{Jiang2024MixtralOE}, and has also been applied to parameter-efficient modules such as LoRA \cite{hu2022lora} during finetuning on downstream tasks \cite{wu2024mole}. We refer readers to the survey paper \cite{cai2024surveymoe} for details on various MoE techniques. Differently from existing works that use MoE to process individual tokens within an LLM, we apply a mixture of audio encoders to process audio samples to input to an LLM.
In vision-language model literature, MoVA \cite{zong2024mova} employs a mixture of vision encoders since no single vision encoder dominates all tasks. However, MoVA needs an additional pass of task instruction prompts through the LLM to obtain routing decisions.

\section{Proposed Method}
\label{sec: proposed method}

\begin{figure}[t]
    \centering
    \begin{subfigure}{\columnwidth}
        \centering
        \includegraphics[width=\linewidth]{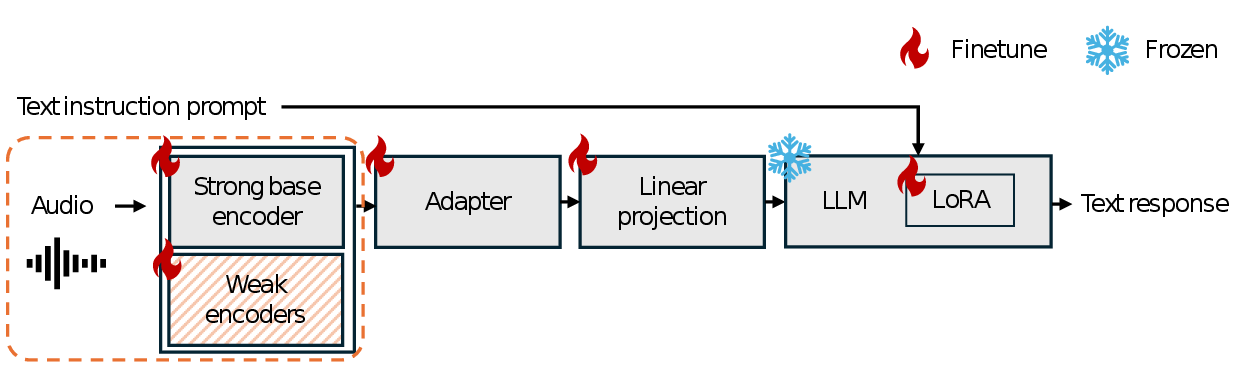}
        \caption{AudioLLM framework} 
        \label{fig: overview framework}
    \end{subfigure}

    \begin{subfigure}{\columnwidth}
        \centering
        \includegraphics[width=\linewidth]{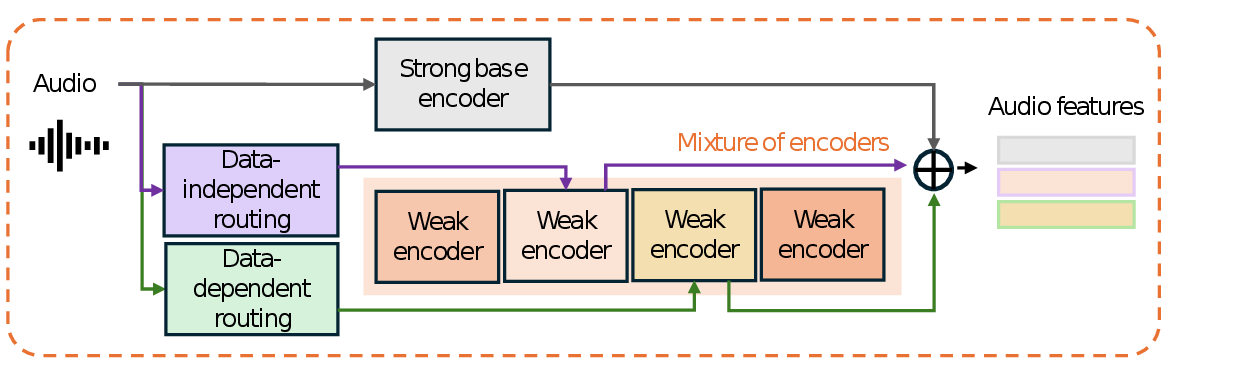}
        \caption{Mixture of weak encoders} 
        \label{fig: overview mowe}
    \end{subfigure}

    \caption{Overview of proposed strategy: (a) AudioLLM framework consists a pool of weak encoders to supplement the strong base encoder in new datasets and tasks, (b) MoWE utilizes a data-independent and a data-dependent router to selectively activate encoders for audio processing.}
    \vspace{-3mm}
\end{figure}

\subsection{AudioLLM Framework}
\label{subsec: framework}

The AudioLLM framework we employ consists (1) a strong base audio encoder $E_{base}$ and $M$ weak audio encoders $\{E_k\}_{k=1}^M$ to be activated by our proposed MoWE strategy, (2) a modality adapter to downsample the encoder embeddings and a linear projection to project the downsampled embeddings into the LLM input space, and (3) an LLM to generate free-form text. The weak encoders are included to supplement the base encoder to learn new datasets and tasks. An overview of the framework is illustrated in Figure~\ref{fig: overview framework}.

We denote the data as $(\mathcal{A}, \mathcal{T}, \mathcal{Y})$, where $\mathcal{A}$ is the input audio, $\mathcal{T}$ is the input text instruction, and $\mathcal{Y}$ is the output text response.
For an audio input $a_i \in \mathcal{A}$ corresponding to the $i$-th sample, the base encoder produces embeddings $z_{i,base} = E_{base}(a_i)$, and a subset of the weak encoders are activated by our proposed MoWE strategy to produce the embeddings $z_{i,MoWE} = MoWE(a_i)$, further described in Section~\ref{subsec: mixture of weak encoders}. The embeddings are concatenated along the feature dimension as $z_i = z_{i,base} \oplus_f z_{i,MoWE}$. Note that the embeddding sequence length and subsequently number of tokens are not increased. We obtain audio tokens $token_{a_i} = proj(adapter(z_i))$, and text instruction tokens $token_{t_i} = tokenizer(t_i)$ from the text instruction $t_i \in \mathcal{T}$. The tokens are concatenated along the sequence dimension as $token_i = token_{a_i} \oplus_s token_{t_i}$, and input into the LLM to generate a text response $\hat{y}_i = LLM(token_i)$. 

During training, we initialize the audio encoders and LLM with pre-trained model weights. The encoders, MoWE routers, adapter, linear projection layer and a light-weight LoRA module \cite{hu2022lora} inserted into the LLM are finetuned. The model is trained with the standard next-token prediction objective $L_{next-token}$ and a MoWE routing loss $L_{MoWE}$:
\begin{equation}
 L = L_{next-token} + 0.1 * L_{MoWE}.
\end{equation}

\subsection{Mixture of Weak Encoders}
\label{subsec: mixture of weak encoders}

Given a pool of $M$ weak encoders $\{E_i\}_{i=1}^M$, as illustrated in Figure~\ref{fig: overview mowe}, we select and activate a subset of the encoders per sample based on (i) a data-independent router, and (ii) a data-dependent router. 

\noindent \textbf{Data-independent routing:} The data-independent router selects a fixed weak encoder irregardless of the input sample, with the purpose of supplementing the overall representation learning capacity of the base encoder. For an audio input $a_i$, 
\begin{align}
    r_{indep} &= KeepTop1(Softmax(w_{indep})) \\
    z_{i,indep} &= \sum_k r_{indep}[k] * E_k(a_i)
\end{align}
where $w_{indep}, KeepTop1(v) \in \mathbb{R}^M$, with $KeepTop1(v)[j]=v[j]$ for $j=argmax(v)$ and $0$ otherwise.

\noindent \textbf{Data-dependent routing:} The data-dependent router selects a weak encoder depending on the input audio to incorporate detailed, data-specific features. For an audio input $a_i$,
\begin{align}
    r_{i,dep} &= KeepTop1(Softmax(\bar{z}_{i,base} W_{dep})) \\
    z_{i,dep} &= \sum_k r_{i,dep}[k] * E_k(a_i)
\end{align}
where $\bar{z}_{i,base} \in \mathbb{R}^{1\times d_{base}}$ is the base encoder embeddings $z_{i,base}$ averaged across the sequence dimension, and $W_{dep} \in \mathbb{R}^{d_{base} \times M}$. The vector $r_{i,dep}$ weigh the contribution of each weak encoder. During training, we smooth the weight $r_{i,dep}$ by $r_{i,dep} = 0.9 * r_{i,dep} + 0.1 * \epsilon$ with $\epsilon = 0.1/M$ to prevent training from being heavily biased towards specific encoders.

Embeddings from the two routers are concatenated across the feature dimension to form $z_{i,MoWE} = z_{i,dep} \oplus_f z_{i,indep}$. The routers are trained alongside other model weights using 
\begin{equation}
    L_{MoWE}= \frac{1}{2} [L_{indep-ent} + (L_{dep-ent} + L_{dep-div})]   
\end{equation}
where
\begin{align}
    L_{indep-ent} &= - r_{indep} * log(r_{indep}) \label{eqn: indep-ent} \\
    L_{dep-ent} &= - \frac{1}{B}\sum_{i=1}^B r_{i,dep} * log(r_{i,dep}) \label{eqn: dep-ent}\\
    L_{dep-div} &= \bar{r}_{dep} * log(\bar{r}_{dep}) \label{eqn: dep-div}
\end{align}
with $\bar{r}_{dep}=\frac{1}{B}\sum_{i=1}^B r_{i,dep}$ being $r_{i,dep}$ averaged across a batch of size $B$.
The entropy losses in Equation~\ref{eqn: indep-ent} and \ref{eqn: dep-ent} encourages confident routing decisions, and the diversity loss in Equation~\ref{eqn: dep-div} prevents data-dependent router decisions from collapsing to a single encoder choice.
\section{Experimental Setup}
\label{sec: experimental setup}

\subsection{Tasks and Datasets}

We demonstrate the effectiveness of our method on a set of 5 popular speech and audio tasks, with one representative dataset for each task as summarized in Table~\ref{tab: data_desc}: LibriSpeech-Clean \cite{panayotov2015librispeech} for Automatic Speech Recognition (ASR), MELD-Emotion \cite{poria2019meld} for Emotion Recognition (ER), Clotho-AQA \cite{lipping2022clothoaqa} for Audio Question Answering (AQA), Spoken-SQuAD \cite{lee2018spokensquad} for Speech Question Answering (SQA), and AudioCaps (AC) \cite{kim2019audiocaps} for Audio Captioning. Audio inputs are trimmed or padded to 30-second length. We perform evaluations with AudioBench \cite{wang2024audiobench}. ASR is evaluated with word error rate (WER), AC is evaluated with METEOR, and the quality of ER, AQA, SQA and AC outputs are graded by Llama-3-70B-Instruct \cite{llama3modelcard} on a 0-5 scale, with a higher score reflecting outputs that more closely resemble the ground-truth.

\begin{table}[t]
\centering

\caption{Datasets for training and evaluation. Tasks included are ASR (Automatic Speech Recognition). ER (Emotion Recognition), AQA (Audio Question Answering), SQA (Speech Question Answering), and AC (Audio Captioning). Model-as-Judge is conducted with Llama-3-70B-Instruct model on a 0-5 grading scale.
\label{tab: data_desc}}

\begin{adjustbox}{max width=\columnwidth}
\begin{tabular}{P{1.5cm}P{2.5cm}P{5cm}}
\toprule[1pt]\midrule[0.3pt]
\textbf{Task} & \textbf{Dataset} & \textbf{Metric} \\ \midrule
ASR & LibriSpeech-Clean & WER (\%) \\
ER  & MELD-Emotion      & Model-as-Judge (0-5 score) \\
AQA & Clotho-AQA        & Model-as-Judge (0-5 score) \\
SQA & Spoken-SQuAD      & Model-as-Judge (0-5 score) \\
AC  & AudioCaps         & METEOR, Model-as-Judge (0-5 score) \\
\midrule[0.3pt]\bottomrule[1pt]
\end{tabular}
\end{adjustbox}


\end{table}

\subsection{Implementation}

In our experiments, we primarily use Whisper-large-v3 \cite{radford2023whisper} with 637M parameters as the strong base encoder, a linear layer plus GELU activation as the adapter to downsample encoder embeddings to 100 tokens, and Llama-3-8B-Instruct \cite{llama3modelcard} as the LLM. We also demonstrate method effectiveness on other choices of LLM such as Zephyr-7B-$\beta$ \cite{tunstall2023zephyr} and a smaller Phi-3-Mini-4K-Instruct \cite{Abdin2024Phi3TR} with 3.8B parameters. In Section~\ref{subsec: mixture of uniform encoders}, we implement MoWE as a mixture of 4 Whisper-tiny encoders, each with 9M parameters. The parameter $w_{indep}$ is randomly initialized using a standard Gaussian distribution.
In Section~\ref{subsec: mixture of diverse encoders} and \ref{subsec: comparison with large scale training}, we use a more diverse mixture of 2 Whisper-tiny and 2 HuBERT encoders, one of which is a HuBERT-base \cite{hsu2021hubert} and the other is a task-specific ER-finetuned encoder \cite{hubert-base-superb-er} (denoted here as HuBERT-base-ER) with 95M parameters each. HuBERT features are linearly interpolated to Whisper-tiny's feature dimensions. We set a prior on the data-independent routing parameter $w_{indep}$ to preferentially select the HuBERT-base-ER i.e. with value 1 for the encoder and -1 otherwise, since ER is a known task in our setup. We conduct multi-task training with batch size 32 for 5 epochs. We used AdamW optimizer with $\beta_1=0.9$ and $\beta_2=0.999$, a learning rate of $5\times 10^{-5}$ with cosine scheduler.
\section{Results and Analysis}
\label{sec: results and analysis}

\subsection{Mixture of Uniform Encoders}
\label{subsec: mixture of uniform encoders}

From Table~\ref{tab: results_1L4T}, MoWE improves the overall model performance on the evaluated tasks across 3 LLMs tested. Although the Whisper-tiny encoders in MoWE have the same initializations and have much smaller size (9M parameters) compared to the strong base encoder (637M parameters), they are still useful in increasing model capacity to learn new tasks.

We perform further analysis on the design of MoWE routers in Table~\ref{tab: results_ablation}. With a single mixture, a data-dependent router achieves better performance than a data-independent router, demonstrating the effect of specialization. With two mixtures, having two data-dependent routers perform suboptimally as they may introduce excessive dynamic changes in the encoder embeddings. A combination of data-dependent router and data-independent router performs the best.
\begin{table}[tb]
\centering

\caption{Performance on each dataset. MoWE comprises a pool of 4 Whisper-tiny encoders. M.J. refers to Model-as-Judge on a 0-5 grading scale.
\label{tab: results_1L4T}}

\begin{adjustbox}{max width=\columnwidth}
\begin{tabular}{cP{1.2cm}*{4}{P{1.0cm}}P{1.65cm}}
\toprule[1pt]\midrule[0.3pt]
\textbf{MoWE} & \textbf{LibriSpeech-Clean} & \textbf{MELD-Emotion} & \textbf{Clotho-AQA} & \textbf{Spoken-SQuAD} & \multicolumn{2}{c}{\textbf{AudioCaps}} \\ \cmidrule(lr){2-2} \cmidrule(lr){3-3} \cmidrule(lr){4-4} \cmidrule(lr){5-5} \cmidrule(lr){6-7}
& WER $(\downarrow)$ & M.J. $(\uparrow)$ & M.J. $(\uparrow)$ & M.J. $(\uparrow)$ & M.J. $(\uparrow)$ & METEOR $(\uparrow)$ \\ \midrule
\multicolumn{7}{c}{Whisper + Llama-3-8B-Instruct} \\ \hdashline
\xmark & 3.17 & 1.41 & 2.91 & 2.82 & 1.93 & 24.42\\
\cmark & \textbf{2.99} & \textbf{1.63} & \textbf{2.99} & \textbf{2.92} & \textbf{1.98} & \textbf{25.70} \\ \midrule
\multicolumn{7}{c}{Whisper + Zephyr-7B-$\beta$} \\ \hdashline
\xmark & 3.29 & \textbf{2.22} & 2.43 & \textbf{3.17} & 1.85 & 26.92 \\
\cmark & \textbf{3.08} & \textbf{2.22} & \textbf{2.51} & 2.90 & \textbf{1.98} & \textbf{28.81} \\ \midrule
\multicolumn{7}{c}{Whisper + Phi-3-Mini-4K-Instruct} \\ \hdashline
\xmark & \textbf{4.28} & \textbf{2.17} & \textbf{1.55} & 2.79 & 1.71 & 26.15 \\
\cmark & 4.35 & \textbf{2.17} & 1.45 & \textbf{2.81} & \textbf{1.76} & \textbf{27.19} \\
\midrule[0.3pt]\bottomrule[1pt]
\end{tabular}
\end{adjustbox}


\end{table}

\begin{table}[tb]
\centering

\caption{Further analysis on MoWE routing. MoWE comprises a pool of 2 Whisper-tiny encoders for 1 mixture, and a pool of 4 Whisper-tiny encoders for 2 mixtures. M.J. refers to Model-as-Judge on a 0-5 grading scale.
\label{tab: results_ablation}}

\begin{adjustbox}{max width=\columnwidth}
\begin{tabular}{P{1.4cm}P{1.2cm}*{4}{P{1.0cm}}P{1.65cm}}
\toprule[1pt]\midrule[0.3pt]
\textbf{MoWE Routing} & \textbf{LibriSpeech-Clean} & \textbf{MELD-Emotion} & \textbf{Clotho-AQA} & \textbf{Spoken-SQuAD} & \multicolumn{2}{c}{\textbf{AudioCaps}} \\ \cmidrule(lr){2-2} \cmidrule(lr){3-3} \cmidrule(lr){4-4} \cmidrule(lr){5-5} \cmidrule(lr){6-7}
& WER $(\downarrow)$ & M.J. $(\uparrow)$ & M.J. $(\uparrow)$ & M.J. $(\uparrow)$ & M.J. $(\uparrow)$ & METEOR $(\uparrow)$ \\ \midrule
\multicolumn{7}{c}{Whisper + 1 Mixture + Llama-3-8B-Instruct} \\ \hdashline
Indep   & 3.03 & 1.13 & 2.83 & 2.92 & \textbf{1.98} & 24.18 \\
Dep     & \textbf{2.93} & \textbf{1.42} & \textbf{2.91} & \textbf{2.94} & 1.97 & \textbf{25.41} \\ \midrule
\multicolumn{7}{c}{Whisper + 2 Mixtures + Llama-3-8B-Instruct} \\ \hdashline
Indep x 2   & \textbf{2.94} & 1.51 & 2.96 & 2.90 & \textbf{2.00} & 25.36 \\
Dep x 2     & 3.13 & 1.15 & 2.71 & 2.82 & 1.98 & 24.61 \\
Indep, Dep  & 2.99 & \textbf{1.63} & \textbf{2.99} & \textbf{2.92} & 1.98 & \textbf{25.70} \\
\midrule[0.3pt]\bottomrule[1pt]
\end{tabular}
\end{adjustbox}


\end{table}

\subsection{Mixture of Diverse Encoders}
\label{subsec: mixture of diverse encoders}

We conducted experiments under a single-stage multi-task training regime and a two-stage training regime where the first stage trains only on the LibriSpeech dataset for ASR to initially align the encoders and LLM, and the second stage trains on all datasets and tasks. Expectedly, the model focuses more on the ASR task with two-stage training. From Table~\ref{tab: results_HB}, we observe that MoWE with a diverse mixture of weak encoders is effective in increasing model performance in both training regimes. In particular, with the inclusion of HuBERT-base-ER in the encoder pool, performance on the ER task is 1.69 and 1.71 under the single-stage and two-stage training regime respectively, higher than the 1.63 achieved in Table~\ref{tab: results_1L4T} where the encoder pool consists of only Whisper-tiny.

\begin{figure}[t]
    \centering
    \includegraphics[width=0.9\columnwidth]{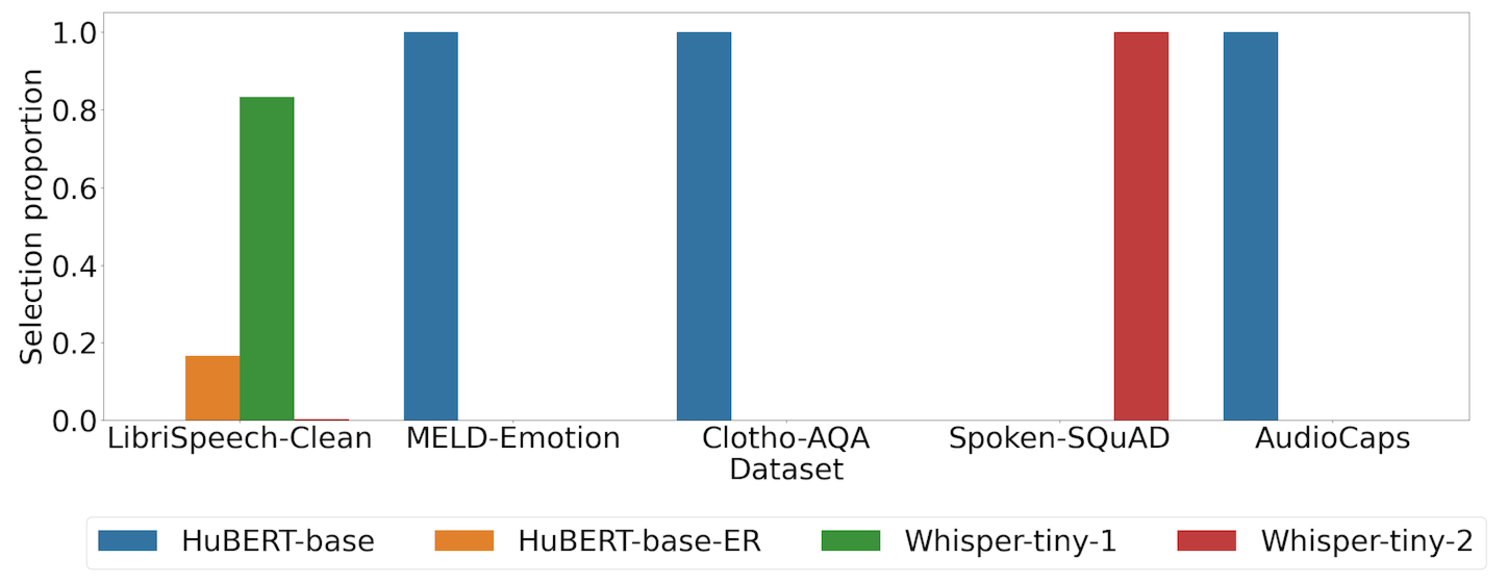}

    \caption{Proportion of samples assigned to each encoder by the data-dependent router. Note that HuBERT-base-ER is activated by the data-independent router for all samples.}
    \label{fig: encoder_selection_proportions}
\end{figure}

In Figure~\ref{fig: encoder_selection_proportions}, for each dataset, we plot the proportion of times each encoder is selected and activated by the data-dependent router during evaluation. Note that since HuBERT-base-ER is already selected by the data-independent router, it is rarely selected again. We observe evident specialization of the encoders. LibriSpeech-Clean and Spoken-SQuAD consist speech mostly in neutral and consistent tone, and are mainly processed by two indiviudal Whisper-tiny encoders that are pre-trained to specialize in ASR. MELD-Emotion consists utterances with varied emotions, and Clotho-AQA and AudioCaps consist non-speech audio, and these datasets are processed by HuBERT-base pre-trained for representation learning by self-supervised learning methods.
\begin{table}[t]
\centering

\caption{Performance on each dataset. MoWE comprises a pool of 2 Whisper-tiny encoders and 2 HuBERT-base encoders (a base and an ER-finetuned encoder). M.J. refers to Model-as-Judge on a 0-5 grading scale.
\label{tab: results_HB}}

\begin{adjustbox}{max width=\columnwidth}
\begin{tabular}{cP{1.2cm}*{4}{P{1.0cm}}P{1.65cm}}
\toprule[1pt]\midrule[0.3pt]
\textbf{MoWE} & \textbf{LibriSpeech-Clean} & \textbf{MELD-Emotion} & \textbf{Clotho-AQA} & \textbf{Spoken-SQuAD} & \multicolumn{2}{c}{\textbf{AudioCaps}} \\ \cmidrule(lr){2-2} \cmidrule(lr){3-3} \cmidrule(lr){4-4} \cmidrule(lr){5-5} \cmidrule(lr){6-7}
& WER $(\downarrow)$ & M.J. $(\uparrow)$ & M.J. $(\uparrow)$ & M.J. $(\uparrow)$ & M.J. $(\uparrow)$ & METEOR $(\uparrow)$ \\ \midrule
\multicolumn{7}{c}{Single-stage training} \\ \hdashline
\xmark & 3.17 & 1.41 & 2.91 & 2.82 & 1.93 & 24.42\\
\cmark & \textbf{3.03} & \textbf{1.69} & \textbf{2.97} & \textbf{2.95} & \textbf{1.97} & \textbf{25.78} \\ \midrule
\multicolumn{7}{c}{Two-stage training} \\ \hdashline
\xmark & \textbf{2.82} & 1.66 & 2.63 & 2.89 & 1.90 & \textbf{24.88} \\
\cmark & 2.85 & \textbf{1.71} & \textbf{2.92} & \textbf{2.93} & \textbf{1.99} & 24.18 \\
\midrule[0.3pt]\bottomrule[1pt]
\end{tabular}
\end{adjustbox}

\vspace{-2mm}

\end{table}

\subsection{Comparison with Models with Large-Scale Training}
\label{subsec: comparison with large scale training}

In Table~\ref{tab: results_comparison}, we verify that that an AudioLLM with MoWE can obtain task performance competitive with state-of-the-art models described in Section~\ref{sec: related works}. We applied SpecAugment~\cite{park2019specaugment} to increase data diversity on the 5 selected datasets and used a Conv1D adapter with kernel and stride size 8 to less rigorously downsample the encoder embeddings. We note that although the state-of-the-art models are trained with larger-scale data, MoWE outperformed in ER, AQA and AC. 

\begin{table}[tb]
\centering

\caption{Comparison with state-of-the-art AudioLLMs. 
\label{tab: results_comparison}}

\begin{adjustbox}{max width=\columnwidth}
\begin{tabular}{cP{1.2cm}*{4}{P{1.0cm}}P{1.65cm}}
\toprule[1pt]\midrule[0.3pt]
\textbf{Model} & \textbf{LibriSpeech-Clean} & \textbf{MELD-Emotion} & \textbf{Clotho-AQA} & \textbf{Spoken-SQuAD} & \multicolumn{2}{c}{\textbf{AudioCaps}} \\ \cmidrule(lr){2-2} \cmidrule(lr){3-3} \cmidrule(lr){4-4} \cmidrule(lr){5-5} \cmidrule(lr){6-7}
& WER $(\downarrow)$ & M.J. $(\uparrow)$ & M.J. $(\uparrow)$ & M.J. $(\uparrow)$ & M.J. $(\uparrow)$ & METEOR $(\uparrow)$ \\ \midrule
SALMONN     & 45.28 & 1.60 & 2.75 & 3.15 & 1.83 & 22.79 \\
Qwen-Audio  & 2.20  & 1.46 & 2.88 & 3.03 & \textbf{2.11} & 22.91 \\
WavLLM      & \textbf{2.05}  & 1.45 & 2.16 & \textbf{3.88} & 0.50 & 6.70 \\
MoWE        & 2.42  & \textbf{1.91} & \textbf{3.14} & 3.13 & 2.07 & \textbf{25.49} \\
\midrule[0.3pt]\bottomrule[1pt]
\end{tabular}
\end{adjustbox}


\end{table}

\section{Further Analysis}

We evaluate MoWE on additional out-of-distribution datasets not seen at training: LibriSpeech-Other \cite{panayotov2015librispeech} for Automatic Speech Recognition, SLUE-P2-SQA \cite{Shon2022SLUEPA} for Speech Question Answering, WavCaps \cite{mei2023wavcaps} for Audio Captioning and WavCaps-AQA \cite{mei2023wavcaps} for Audio Question Answering. The datasets are processed and the evaluation is performed with AudioBench \cite{wang2024audiobench}. From Table~\ref{tab: results_ood}, we note that MoWE performs comparatively with the baseline on these out-of-distribution datasets. This is not surprising since (multimodal) LLM capabilities depend on the training data and improving generalization is still an open problem.
\begin{table}[tb]
\centering

\caption{Performance on out-of-distribution datasets not seen at training. M.J. refers to Model-as-Judge on a 0-5 grading scale.
\label{tab: results_ood}}

\begin{adjustbox}{max width=\columnwidth}
\begin{tabular}{c*{4}{P{1.8cm}}}
\toprule[1pt]\midrule[0.3pt]
\textbf{MoWE} & \textbf{LibriSpeech-Other} & \textbf{SLUE-P2-SQA} & \textbf{WavCaps} & \textbf{WavCaps-AQA} \\ \cmidrule(lr){2-2} \cmidrule(lr){3-3} \cmidrule(lr){4-4} \cmidrule(lr){5-5}
& WER $(\downarrow)$ & M.J. $(\uparrow)$ & M.J. $(\uparrow)$ & M.J. $(\uparrow)$ \\ \midrule
\xmark & 5.57 & \textbf{3.85} & \textbf{1.19} & 1.37 \\
\cmark & \textbf{5.50} & 3.82 & 1.16 & \textbf{1.44} \\
\midrule[0.3pt]\bottomrule[1pt]
\end{tabular}
\end{adjustbox}


\end{table}

We provide details on the total number of parameters, number of active parameters, and throughput measured by number of samples processed per second in Table~\ref{tab: efficiency}. For the count of active parameters, we assume samples are drawn from each dataset with equal probability. Throughput is computed on one Nvidia H100 GPU. Adding the mixture of weak encoders only slightly reduces throughput, as the parameter counts of the weak encoders are significantly smaller in comparison with the LLM (9-95M vs. 8B).
\begin{table}[tb]
\centering

\caption{Computational efficiency at inference time.
\label{tab: efficiency}}

\begin{adjustbox}{max width=0.8\columnwidth}
\begin{tabular}{c*{3}{P{1.8cm}}}
\toprule[1pt]\midrule[0.3pt]
\textbf{MoWE} & \textbf{\# total parameters} & \textbf{\# active parameters} & \textbf{\# samples per second} \\ \midrule
\xmark & 8.73B & 8.73B & 1.92\\
\cmark & 8.96B & 8.91B & 1.83 \\
\midrule[0.3pt]\bottomrule[1pt]
\end{tabular}
\end{adjustbox}


\end{table}

\section{Conclusion}
\label{sec: conclusion}

In this paper, we proposed MoWE, a novel approach that integrates a mixture of weak encoders into the AudioLLM framework. MoWE supplements the base encoder by expanding its feature extraction capacity and capabilities without significantly increasing model size. By incorporating a diverse set of encoders, the approach encourages encoder specialization and the learning of data-specific features. Moreover, MoWE demonstrated improved performance in multi-task settings and achieves competitive results when compared to state-of-the-art models.

\section*{Acknowledgements}
This research is supported by the National Research Foundation, Singapore under its National Large Language Models Funding Initiative. Any opinions, findings, conclusions, or recommendations expressed in this material are those of the author(s) and do not reflect the views of National Research Foundation, Singapore.


\bibliographystyle{IEEEbib}
\bibliography{references}    

\end{document}